\documentclass[%
 aip,
 amsmath,amssymb,
reprint,%
]{revtex4-2}

\usepackage{graphicx}
\usepackage{dcolumn}
\usepackage{lipsum}
\usepackage{bm}
\usepackage{tabu}
\usepackage[utf8]{inputenc}
\usepackage[T1]{fontenc}
\usepackage{mathptmx}
\usepackage{color}

\usepackage{tikz}
\newcommand\encircle[1]{%
  \tikz[baseline=(X.base)] 
    \node (X) [draw, shape=circle, inner sep=-.5] {\strut #1};}
\begin{document}
\preprint{AIP/123-QED}
\title[Generation of multiple vector beams through cascaded beam displacers and a segmented digital hologram]{Generation of multiple vector beams through cascaded beam displacers and a segmented digital hologram}
\author{Bo-Zhao}
\affiliation{Wang Da-Heng Collaborative Innovation Center for Quantum manipulation \& Control, Harbin University of Science and Technology, Harbin 150080, China}%
\author{Jia-Yuan Wu}
\affiliation{Department of Applied Physics, School of Science, Harbin University of Science and Technology, Harbin, 150080, China}%
\author{Xiang-Yu Yu}
\affiliation{Department of Applied Physics, School of Science, Harbin University of Science and Technology, Harbin, 150080, China}%
\author{Xiao-Bo Hu}
\email{huxiaobo@zstu.edu.cn}
\affiliation{Key Laboratory of Optical Field Manipulation of Zhejiang Province,Department of Physics, Zhejiang Sci-Tech University, Hangzhou, 310018, China}%
\author{Carmelo Rosales-Guzm\'an}
\email{carmelorosalesg@gmail.com}
\affiliation{Centro de Investigaciones en Óptica, A.C., Loma del Bosque 115, Colonia Lomas del campestre, 37150 León, Gto., Mexico}%

\date{\today} 

\begin{abstract}
Complex vector light modes, characterized by a non-uniform transverse polarization distribution, have pervaded a wide range of research fields. In this study, we propose a novel approach that enables the simultaneous generation of multiple vector beams based on a spatially-segmented digital hologram and two or more cascaded beam displacers. More precisely, an input beam is separated into multiple parallel copies spatially separated, which are then sent to the center of each segmented hologram, enabling independent modulation of each beam. The modulated beams are then judiciously recombined with a beam displacer to generate multiple vector modes in a simultaneous way. We demonstrated our technique with two arbitrary vector modes but the technique can be easily extended to more by inserting additional beam dispalcers. To assess the quality of the generated vector modes, we employed Stokes polarimetry to reconstruct their transverse polarisation distribution and to measure their degree of non-separability. We envision that this technique will find significant applications in various fields, including optical communications, optical sensing, optical tweezers to mention a few.
\end{abstract}

\maketitle
The field of structured light, which involves the manipulation of some of the properties of light to generate novel states of light is completely reshaping the field of modern optics at both, the fundamental and application levels \cite{Roadmap,Shen2022}. Of particular interest and of relevance for this research article are the so-called complex vector beams, nonseparable in their spatial and polarisation degrees of freedom \cite{Zhan2009,Rosales2018Review}. Such beams are paving applications in fields as diverse as optical tweezers, optical communications, optical metrology, amongst others \cite{Yuanjietweezers2021,Ndagano2018}. As such, several techniques for the generation of vector beams with arbitrary polarisation and spatial distributions have been proposed. Amongst all of this, the ones that employees liquid crystal Spatial Light Modulators (SLMs) and Digital Micromirror Devices (DMDs) have gain popularity, in part due to their high speed an flexibility in the generation of arbitrary vector beams at the click of a button \cite{SPIEbook,Mitchell2016,Ren2015,Gong2013,Hu2022}. Here the variety of techniques include the use of interferometric arrays with one or two SLM \cite{Perez-Garcia2017,Perez-Garcia2022} or by passing an input beam two times over a single SLM \cite{Rong2014,Otte2018b,moreno2014,Tang2020,Zhang2023}. Alternative techniques exploit the polarisation-independent property of DMDs\cite{Rosales2020,Liyao2020,Hu2022}. Some others incorporate additional optical elements, such as beam displacers, Wallastone prisms or right angle prism mirrors \cite{Maurer2007,PengLi2018,ShengLiu2018}.

Importantly, for a wide variety of applications, it is highly desired to develop techniques capable to generate multiple vector beams in a simultaneous way, for example in optical tweezers to trap multiple particles in a simultaneous way \cite{Bhebhe2018} in optical communications  to increase the the speed or security of the information transfer \cite{Ndagano2018,Milione2015e,Otte2020}. Along this line, pioneering techniques have proposed a multiplexing approach, which relies in the superposition principle, whereby several holograms are added (multiplexed) onto a single one. Each hologram incorporates a unique carrier frequency to generate multiple scalar beams at different diffraction angles, which are later recombined by interferometric means to generate multiple vector beams\cite{Bhebhe2018a}. One of the main drawbacks of this technique is precisely its interferometric nature, which causes a random fluctuation of the intermodal phase, that ultimately translates into random fluctuations of the vector beam's polarisation distribution.

In this manuscript, we propose a novel method for generating multiple vector light modes based on a segmented digital hologram displayed on a SLM and a set of Beam Displacers (BDs). For its implementation, the hologram displayed on the SLM is divided into four or more equally-spaced sections, depending on the number of desired vector beams, where independent holograms are displayed. In general, the number of required holograms is two times the number of desired vector beams. More precisely, our technique consist on splitting an input beam into multiple copies of itself, this can be achieved by the appropriate combination of beam displacers and Half-Wave plates (HWPs). Here, the number of beams scales as a power of two with the number of beam displacers. Each beams is then directed to the center of the independent hologram, where they acquire specific spatial modulation. Afterwards they are judiciously recombined in pairs, with the help of another BD and a HWP, to generate the number of desired vector modes. By way of example, with three beam displacers we can generate eight input beams, from which four vector beams can be generated. To demonstrate our technique, we generated two arbitrary vector beams by splitting an input beam into four, exactly the same number of independent holograms in which the screen of the SLM is segmented. To assess the quality of the generated beams, We reconstructed their transverse polarization distribution and measured their beam quality factor, by means of Stokes polarimetry, and compared these with numerical simulations. Further, since each beam is modulated in an independent section of the SLM, the quality of the generated vector beam is higher when compared the techniques that rely on spatial multiplexing.

\begin{figure}[t]
    \centering
    \includegraphics[width=.48\textwidth]{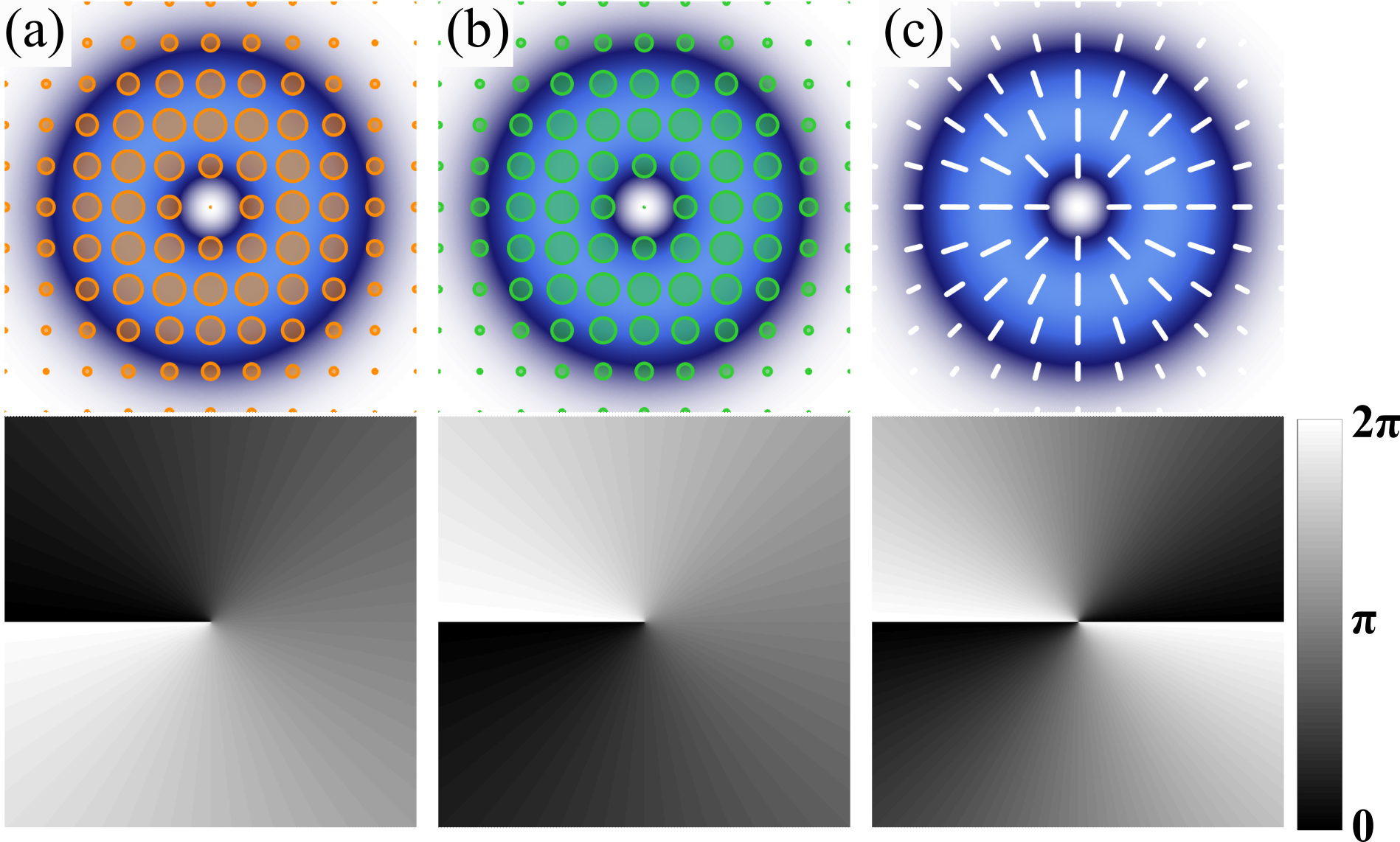}
    \caption{ The top panels show the transverse polarisation distribution overlapped with the intensity profile of  the modes (a) $LG_{0}^{1}$ and (b) $LG_{0}^{-1}$ with right- and left-handed circular polarisation, respectively, to generate the radially-polarized cylindrical vector mode shown in (b). The bottom panels show the corresponding phase distribution of the scalar modes, as well as the intermodal phase of the vector mode.}
    \label{LGvector}
\end{figure}

To begin with, lets recall that complex vector light modes are mathematically described by a nonseparable weighted superposition of the spatial and polarisation degrees of freedom as,
\begin{equation}
    \centering
    {\bf U}=\cos\theta {\bf U}_{1}\hat{\bf e}_r+\sin\theta\exp(i\alpha) {\bf U}_{2}\hat{\bf e}_l,
    \label{Eq:VM}
\end{equation}
where the  coefficients $\cos(\theta)$ and $\sin(\theta)$, with $\theta\in[0,\pi/2]$, are weighting factors that allow a smooth transition of the generated field {\bf U}, from scalar ($\theta=0$ and $\theta=\pi/2$) to vector ($\theta=\pi/4$).  The unitary vectors $\hat{\bf e}_r$ and $\hat{\bf e}_l$, represent the right- and left-handed circular polarization states, respectively, while the term ${\text e}^{i\alpha}$, with $\alpha\in[0,\pi]$ sets a phase difference between both polarization components. In addition, the functions ${\bf U}_{1}$ and ${\bf U}_{2}$ represent orthogonal spatial modes, which can be taken from solutions to the wave equations in either the cylindrical, elliptical parabolic or other coordinate systems. Amongst these, Laguerre-Gaussian ($LG_p^\ell$) modes featuring azimuthally-varying phase of the form $\exp{(i\ell\varphi)}$, is the natural solution to the paraxial wave equation of cylindrical coordinates. The index $p \in \mathbb{N}$ enables the generation of ($p+1$) intensity rings along the radial direction, while the index $\ell \in \mathbb{Z}$, known as the topological charge, is associated to the number of times the phase wraps around the optical axis. By way of example,  Fig.\ref{LGvector} shows the generation process of a Cylindrical Vector Vortex (CVV) beam with radial polarization distribution. For this specific example we used a superposition of the modes $LG_{0}^{1}$ with right circular polarization (see Fig.\ref{LGvector} (a)) and $LG_{0}^{-1}$ with left circular polarization (see Fig.\ref{LGvector} (b)) and parameters $\theta=\pi/4$ and $\alpha=0$. The generated vector beam is shown in Fig. \ref{LGvector} (c). Here, the top panels show their transverse intensity profile, overlapped with their polarization distribution, where right and left circular polarization are represented with orange and green ellipses, respectively, while linear polarization by white lines. The bottom panels show their corresponding phase distribution.

\begin{figure*}[tb]
    \centering
    \includegraphics[width=.8\textwidth]{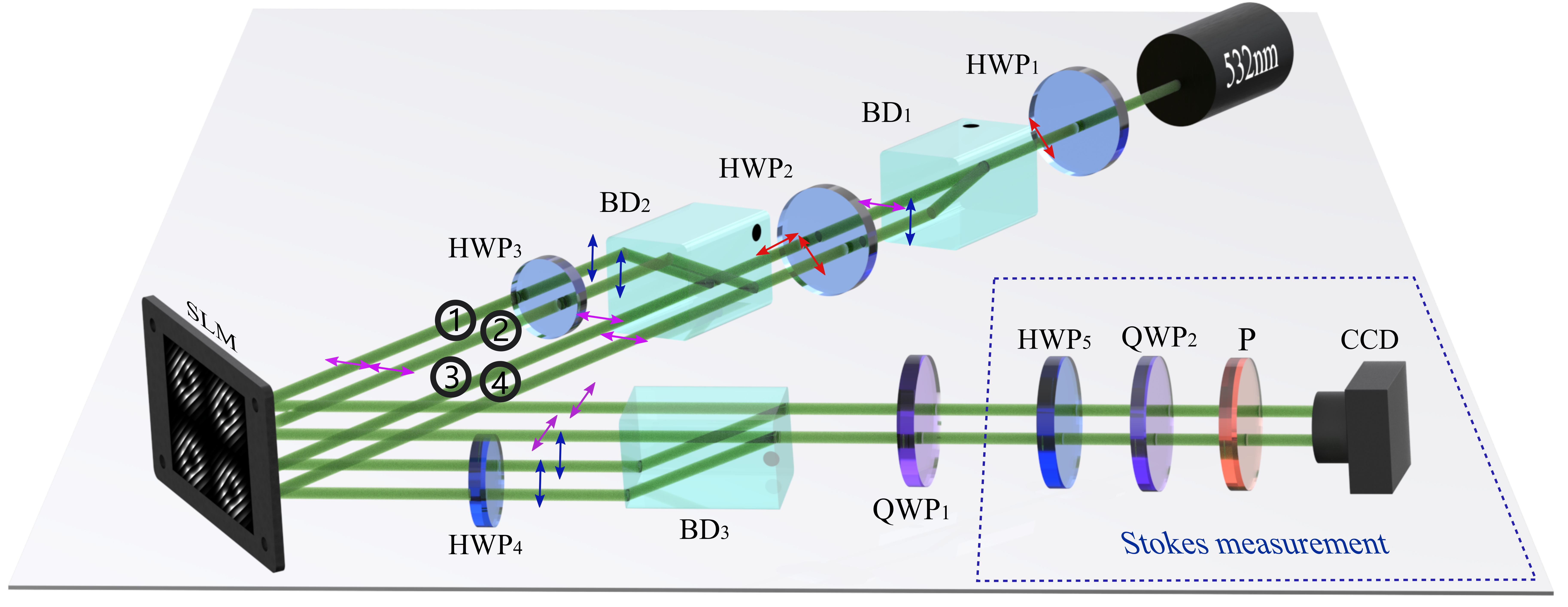}
    \caption{Experimental setup for the simultaneous generation of two complex vector modes. An expanded diagonally-polarised laser beam ($\lambda=532$nm), obtained with the help of a Half-Wave plate (HWP$_1$), is split into its vertical and horizontal polarisation components by means of a Beam Displacer (BD$_1$). Afterwards, HWP$_2$ rotates their polarization to diagonal and antidiagonal, respectively, each of which is split again, by means of BD$_2$, into two with horizontal polarization, and two with vertical. Afterwards HWP$_3$ rotates the beams with vertical polarization to horizontal to match the requirements of the Spatial Light Modulator (SLM). All the beams are then redirected to the SLM, which displays four spatially separated and independent holograms. Finally, BD$_3$ in combination with HWP$_4$ recombines the beams to generate two arbitrry Vector beams. A Quater-Wave plate (QWP) transforms the polarisation basis from linear to circular. Their transverse polarisation distribution is then reconstructed via Stokes polarimetry using a Charged-Coupled Device (CCD) camera.}
    \label{setup}
\end{figure*}
\begin{figure}[tb]
    \centering
    \includegraphics[width=0.45\textwidth]{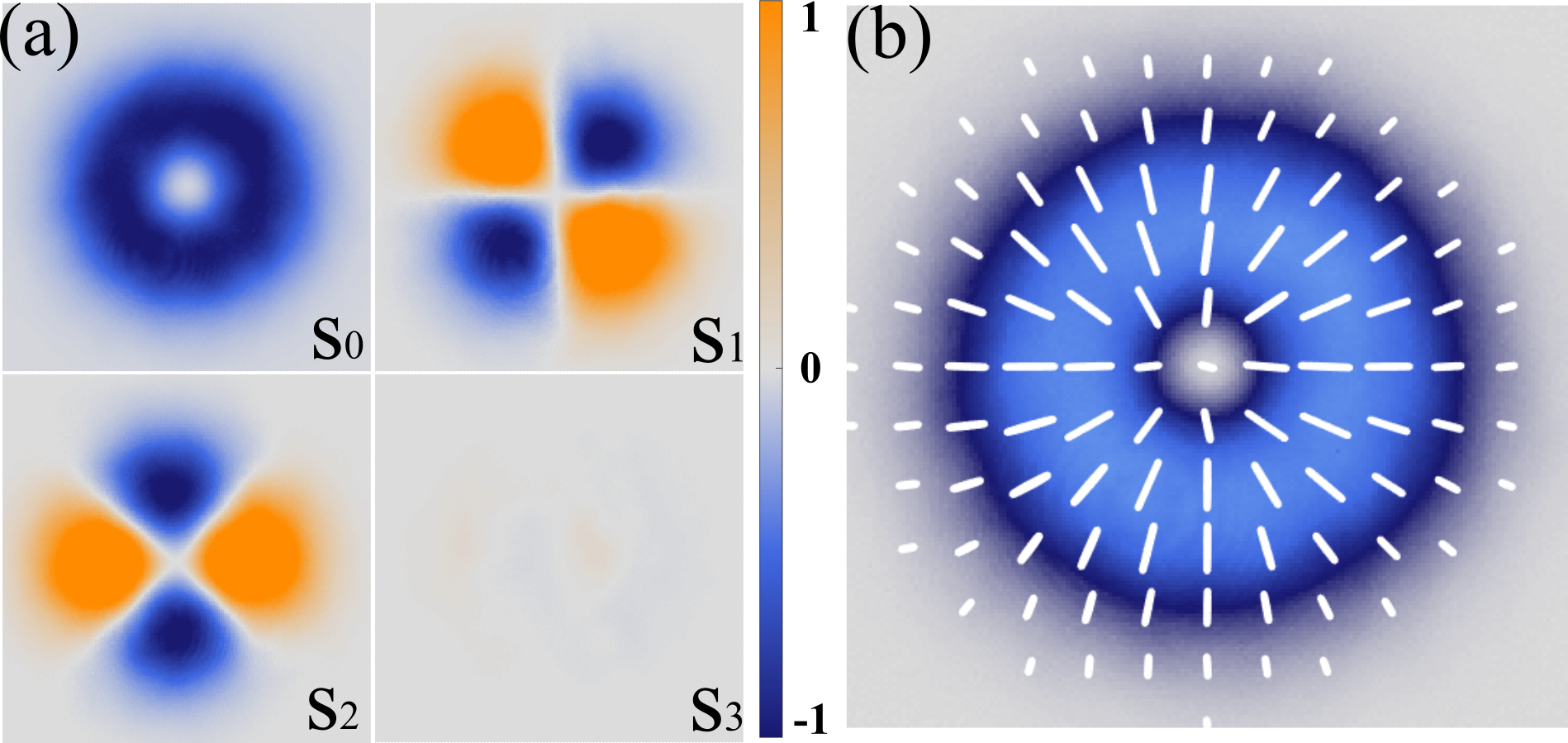}
    \caption{(a) Experimentally measured Stokes parameters $S_0$, $S_1$, $S_2$ and $S_3$ to reconstruct the transverse polarisation distribution of the radially-polarised vector mode shown in (b).}
    \label{Fig:stokes}
\end{figure}

 Our technique takes advantage of a set of Beam Displacers (BDs) and one SLM digitally devided into several segments. Some BDs  are inserted before the hologram which is encoded on the SLM, to split one single input beam exponentially.  Another BD is placed after the SLM to recombine all the modulated beams in pairs. It is worth noting that the number of generated vector modes ($M$) depends on the number of BDs ($N$) according to the mathematical relation $M=2^{(N-2)}$, but also on the size of the BDs and the beam width of the desired beams. A schematic representation of the setup implemented to demonstrate this technique is scheduled in Fig. \ref{setup}. a horizontally polarized laser beam ($\lambda=532$nm) is transformed into a diagonally polarized beam (indicated by the red arrow) using a Half-Wave plate (HWP$_1$) at 22.5$^{\circ}$. The first Beam Displacer (BD$_1$) separates the beam into its horizontal and vertical polarization components, indicated by the pink and blue arrows respectively, at a certain spacing distance. Both beams are then directed in parallel to a second HWP$_2$ at 22.5$^{\circ}$, where one beam is transformed into a diagonal polarization state and the other into an anti-diagonal polarization state. A second Beam Displacer (BD$_2$) then separates both beams into four beams, with two beams \encircle{1} and \encircle{2} having vertical polarization and the other two  \encircle{3} and \encircle{4} having horizontal polarization. Since the SLM can only modulate the horizontal polarization component of light, HWP$_3$ at 45$^{\circ}$ is inserted into the paths of beams to transform their polarization states into horizontal. The SLM is encoded with a digitally-segmented hologram divided into four sub-parts, each allowing independent manipulation of the input beams. The center position of each sub-hologram can be digitally adjusted to match uniform spacing between each beam. After passing through the center of each sub-hologram and being manipulated in their spatial structures, the four beams are recombined using a last BD to generate our object vector modes. Before this, HWP$_4$ at 45$^{\circ}$ is inserted into two of the four paths \encircle{3} and \encircle{4} to transform their polarization states back to vertical. To change our generated vector modes from the linear to the circular polarization basis, we add a Quarter-Wave plate (QWP$_1$) oriented at 45$^{\circ}$ along the path. The transverse polarization was reconstructed by the Stokes polarimetry according to
\begin{equation}\label{Eq:Stokes}
\begin{split}
\centering
     &S_{0}=I_{0},\hspace{19mm} S_{1}=2I_{H}-S_{0},\hspace{1mm}\\
     &S_{2}=2I_{D}-S_{0},\hspace{10mm} S_{3}=2I_{R}-S_{0},
\end{split}
\end{equation}
where $I_0$ is the total intensity of the mode and $I_H$, $I_D$ and $I_R$ the intensity of the horizontal, diagonal and right-handed polarisation components, respectively. The intensities were acquired through the combination of  HWP$_5$,  a QWP$_2$ and a linear polariser (P) and recorded with a Charge-Coupled Device camera (CCD; FL3-U3-120S3C-C with a reslution of 1.55$\mu$m), as detailed in\cite{Zhaobo2019}. $I_H$ was measured by passing the generated two vector fields through a linear polariser at $0^\circ$,  $I_D$ and $I_R$ were measured by passing the beam through an HWP$_5$ at $22.5^\circ$ and  a QWP at $45^\circ$, respectively in front of a polariser at $0^\circ$. Figure \ref{Fig:stokes} shows an example of such measurements for the specific CVV mode with radial polarization distribution. Here, the Stokes parameters $S_{0}$, $S_{1}$, $S_{2}$ and $S_{3}$ are shown in Fig. \ref{Fig:stokes}(a), whereas the corresponding intensity profile overlapped with the polarisation distribution in Fig. \ref{Fig:stokes} (b).

\begin{figure}[tb]
    \centering
    \includegraphics[width=.48\textwidth]{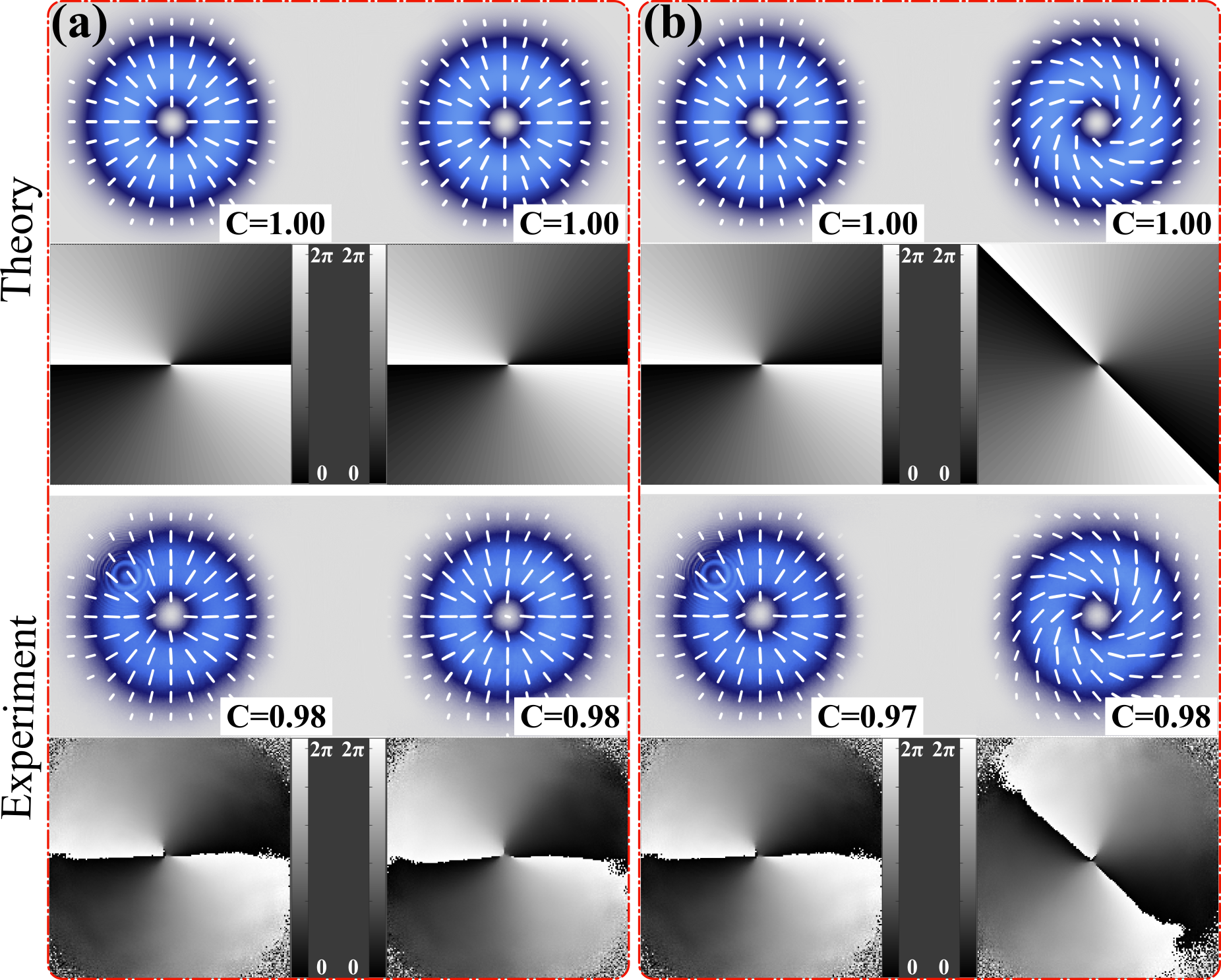}
    \caption{Polarisation distribution overlapped with the intensity profile (top) and intermodal phase profile (bottom) of the vector mode generated through a superposition of the modes (a) $\left\{LG_0^{+1}, LG_{0}^{-1}\right\}$ with parameters $\theta=\pi/4$ and $\alpha=0$. (b) One of the modes is used as reference, while the other is sent through a half-wave plate oriented at $22.5^\circ$. Theoretical and experimental results are shown on the top and bottom, respectively.}
    \label{twins}
\end{figure}

With the setup described above we first generated two identical CVV modes with radial polarization. As a specific example, we employed the combinations of $\left\{LG_0^{+1},LG_{0}^{-1}\right\}$ to generate CVV modes, as illustrated in Fig.\ref{twins}(a), theoretical results on the top panel while experimental on the bottom. In each case, we show the reconstructed polarization distribution from Stokes parameters on the top and the phase of complex Stokes field obtained from $\alpha = \arctan (S_{2}/ S_{1})$ on the bottom. To quantitatively assess the deviation between the experimentally generated modes and the theoretical predictions, we also evaluated the concurrence ($C$), which serves as a measure of the degree of nonseparability between the two degrees of freedom (DoFs) in complex vector mode. A concurrence value of 0 indicates complete decoupling, while a value of 1 complete coupling between the two DoFs. Specifically, $C$ can be determined by comparing the transverse polarization distribution across the entire transverse plane, following the relationship
\begin{equation}
C=\sqrt{1-\left(\frac{\mathbb{S}_1}{\mathbb{S}_0} \right)^2-\left(\frac{\mathbb{S}_2}{\mathbb{S}_0} \right)^2-\left(\frac{\mathbb{S}_3}{\mathbb{S}_0} \right)^2},
\end{equation}
where, $\mathbb{S}_i$ are the values of Stokes parameters $S_i$ integrated over the entire transverse profile. The calculated concurrence values $C$  are shown as insets in the middle of each panel, where $C=1$ for the theoretical cases and $C=0.98$ for both experimental cases. These values exemplify the high performance of our technique to generate two vector beams simultaneously, compared to the numerically-simulated vector modes. 

A possible application of the simultaneous generation of two vector beams is in the field of optical metrology, for example, to determine a specific property of a medium, lets say its temperature \cite{Zhaobo2019}, it is customary to interrogate it with a single beam. The temperature is then determined by correlating the beam with and without the medium, which does not allow real-time applications in swiftly altering surroundings. Hence, the proposed technique, which allows the simultaneous generation of two (or more) vector beams, empowers the exploration of vast potential applications, whereby one of the two generated beams serves as the reference. By way of example, we generated two identical vector modes through the superposition of the modes $\left\{LG_0^{+1}, LG_{0}^{-1}\right\}$, with parameters $\theta=\pi/4$ and $\alpha=0$. One of the modes acts as the reference beam, while the other is used as the interrogating beam and sent through a HWP at an angle of $22.5^\circ$ rotating the entire polarisation state by $45^\circ$ as shown in Fig. \ref{twins}(b). For the sake of comparison, the theoretical and experimental outcomes are displayed in the upper and lower panels of Fig. \ref{twins}(b), respectively.

\begin{figure}[ht]
    \centering
    \includegraphics[width=0.48\textwidth]{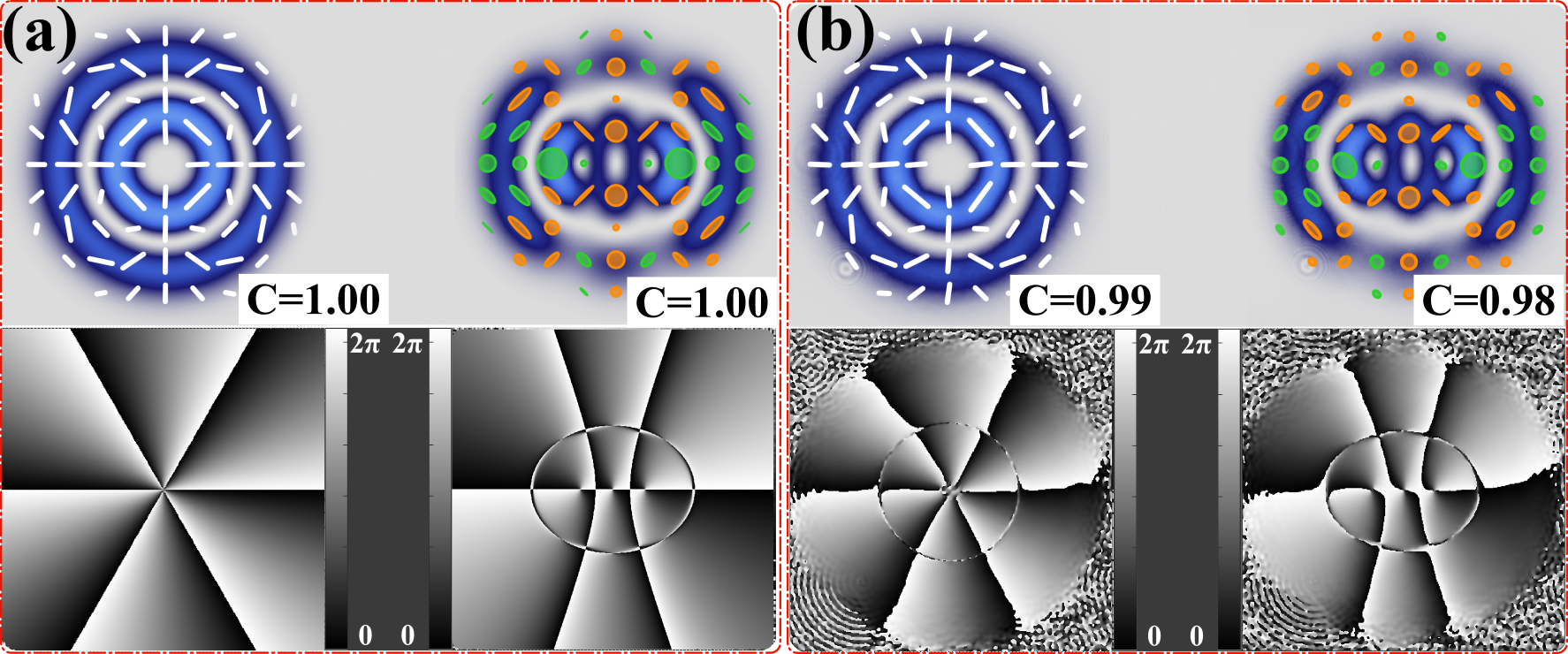}
    \caption{Generation of two different types of vector modes. Theoretical (a) and experimental (b) transverse intensity profile overlapped with the polarisation distribution on the top, while the intromodal phase profile on the bottom. Here we show the specific case for (a) one vector mode with scalar basis of $\left\{LG_1^{-3}, LG_{1}^{+3}\right\}$ and the other with  $\left\{IG_{5,3,2}^{e}, IG_{5,3,2}^{o}\right\}$.}
    \label{Different}
\end{figure}

As a final example and to show the versatility of our technique, we show a representative example of two vector modes generated in different spatial basis, cylindrical and elliptical, respectively. The first case, which is illustrated on the left of Fig. \ref{Different}(a) and (b), theory and experiment, respectively, corresponds to the CVV mode generated with the scalar modes $\left\{LG_1^{+3}\right\}$ and $\left\{LG_{1}^{-3}\right\}$ with parameters $\theta=\pi/4$ and $\alpha=0$. The second case, shown on the right side of Fig. \ref{Different}(a) and \ref{Different}(a), theory and experiment, corresponds to an Ince-Gaussian vector mode given by\cite{Liyao2020},
\begin{equation}
    \Psi_{p,m,\varepsilon}({\bf r}) = \cos\theta IG_{p,m,\varepsilon}^e({\bf r}){\bf e}_r + \sin\theta \exp(i\alpha) IG_{p,m,\varepsilon}^o({\bf r}){\bf e}_l,    
\end{equation}\label{InceModes}
where, $IG_{p,m,\varepsilon}^e$ and $IG_{p,m,\varepsilon}^o$ are the scalar even and odd Ince-Gaussian modes \cite{Bandres2004}. Further, the indexes $p,m\in\mathbb{N}$ obey the relations $0\leq m\leq p$ for even modes and $1\leq m\leq p$ for odd, where as $\varepsilon\in[0.\infty)$ is the ellipticity of the mode. In particular, for this example we used the parameters $p=5$, $m=3$ and $\varepsilon=2$. In each figure, the overlapping transverse polarization distribution with the intensity profile are depicted on the top while the phase of complex Stokes field on the bottom. Additionally, the concurrence ($C$) values for each case are computed and displayed as insets in each panel. As expected, $C$ remains close to one for all cases.

In summary, in this manuscript we introduced a novel technique capable of generating multiple complex vector modes simultaneously. Our approach maximizes the utility of a single digital hologram, which was segmented into multiple sections, enabling high-resolution independent modulation of each scalar beam. To split and recombine the beams, a certain number of beam displacers are required, in direct correlation with the quantity of required vector modes. Significantly, our device boasts compactness and straightforward implementation, yet it remains exceptionally powerful. This capability allows for the generation of multiple arbitrary complex vector modes without limitations on their spatial and polarization distribution. In this manuscript, we showcase the application of this technique in generating two arbitrary vector modes. Through Stokes polarimetry, we reconstruct the transverse polarization distribution and conduct theoretical as well as experimental analyses of the concurrence for each case. Our experimental results exhibit a strong alignment with theoretical predictions, further underscoring the robustness of our technique. We envision that our approach will pave the way for real-world applications in optical sensing, optical communications, optical tweezers, imaging, to mention a few.\\
\noindent

{\bf DATA AVAILABILITY}\\
The data that support the findings of this study are available from the corresponding author upon reasonable request.\\
This work was partially supported by the National Natural Science Foundation of China (61975047),   Natural Science Foundation of Heilongjiang Province (LH2022A016), and Zhejiang Provincial Natural Science Foundation of China (LQ23A040012).
\vspace{-.8cm}

\section*{References}
%

\end{document}